\documentclass[10pt]{article}

\usepackage[a4paper,top=20mm,bottom=18mm,left=27mm,right=27mm]{geometry}
\usepackage[T1]{fontenc}
\usepackage{lmodern}
\usepackage{microtype}
\usepackage{amsmath,mathtools}
\usepackage{booktabs,tabularx,array}
\usepackage{enumitem}
\usepackage{graphicx}
\usepackage{xcolor}
\usepackage{tikz}
\usetikzlibrary{arrows.meta,calc,decorations.pathreplacing,fit,positioning}
\usepackage{placeins}
\usepackage{float}
\usepackage[skip=6pt plus 1pt,indent=0pt]{parskip}
\usepackage[numbers,sort&compress]{natbib}
\usepackage[hidelinks]{hyperref}

\setlist[itemize]{leftmargin=1.4em,itemsep=0.15em,topsep=0.35em}
\newcommand{\modelname}{TontaubeV1}
\newcommand{\cb}[1]{\ensuremath{\mathrm{CB}_{#1}}}

\definecolor{PromptFill}{HTML}{E9E7FF}
\definecolor{PromptStroke}{HTML}{625DB5}
\definecolor{TextFill}{HTML}{E7F2F8}
\definecolor{TextStroke}{HTML}{4D7896}
\definecolor{MarkerFill}{HTML}{C6ECF4}
\definecolor{MarkerStroke}{HTML}{27869A}
\definecolor{AudioFill}{HTML}{FFF0DC}
\definecolor{AudioStroke}{HTML}{B97536}
\definecolor{TargetFill}{HTML}{F8E1EB}
\definecolor{TargetStroke}{HTML}{A64F77}
\definecolor{PadFill}{HTML}{ECEFF1}
\definecolor{PadStroke}{HTML}{7C878F}
\definecolor{DiagramPath}{HTML}{384D5C}

\definecolor{SepColor}{HTML}{C0392B}
\newcommand{\sep}{\,\textcolor{SepColor}{\scriptsize\textperiodcentered}\,}

\title{\bfseries TontaubeV1: Streaming Text-to-Speech\\
with Hierarchical Codec Modeling\\
and Bounded Context}

\author{Fritz Cremer\thanks{These authors contributed equally to this work.} \hspace{2.5em} Jonathan Cremer\footnotemark[1]\\[6pt]
\small \raisebox{-0.25em}{\includegraphics[height=1.2em]{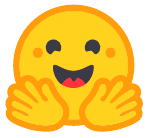}}\hspace{0.5em}%
\url{https://huggingface.co/TontaubeAI/TontaubeV1}\\[3pt]
\small \raisebox{-0.25em}{\includegraphics[height=1.2em]{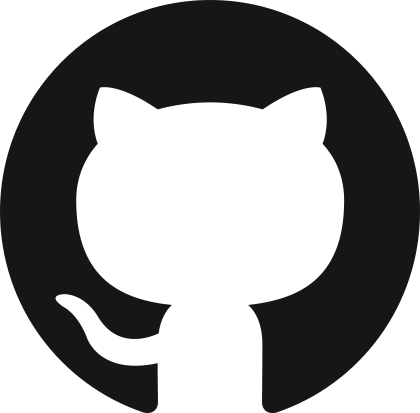}}\hspace{0.5em}%
\url{https://github.com/craitech/tontaube}}
\date{}

\begin{document}
\maketitle
\vspace{6pt}

\begin{abstract}
\noindent
Text-to-speech systems often face a trade-off between natural prosody and efficient inference: higher perceptual quality usually comes at increased computational cost and latency. We present \modelname{}, a model that preserves natural prosody while enabling streaming from a single consumer GPU. Speech is encoded by the hierarchical
DualCodec representation at 12.5\,Hz, which separates a semantic stream from
successive acoustic refinements. Our design assumes that prosodic structure is
largely established when the semantic stream is generated, and allocates
capacity accordingly: a Qwen3-1.7B-derived transformer predicts that stream and
thereby the utterance duration, while three progressively smaller
Qwen3-0.6B-derived transformers each add one acoustic refinement.
Text is tokenized per character rather than by subword. Paired text
and audio markers at shared positions support long-form generation
with bounded context, and overlapping DualCodec reconstructions are mapped into
the VibeVoice acoustic latent space and decoded causally, enabling streaming
despite DualCodec's noncausal decoder. The model
accepts up to one minute of reference audio for voice conditioning and is
designed primarily for English and German, with additional multilingual
support. The four predictors total 2.9B parameters; on a single RTX 5090 the
streaming path reaches about 200\,ms to first audio. In separate
non-streaming measurements, the end-to-end real-time factor (RTF) is 0.08 for
one input and the aggregate RTF is 0.02 across eight concurrent inputs. In a
blinded study, 40 listeners preferred \modelname{}'s prosody over Fish Audio S2
Pro in 70.25\% of 400 comparisons (95\% CI: 64.2--75.7\%). On an
LLM-as-a-judge benchmark, it matches ElevenLabs Flash v2.5 and outperforms the
April 2026 Gradium API and Cartesia Sonic 3 on prosody. The weights are
released on
Hugging Face under the Tontaube Community Model License 1.0.
\end{abstract}

\section{Introduction}
\label{sec:introduction}

Serving text-to-speech at low cost while maintaining high quality presents a
trade-off. Larger systems generally produce more natural speech but occupy more
GPU memory, cost more per hour of audio, and often take longer to
produce their first audio; smaller systems are cheaper and faster but tend to
produce flatter prosody, place emphasis poorly, or lose prosodic consistency over longer
passages. \modelname{} targets the lower-cost end of this range while achieving
competitive prosodic quality in our audiobook-reading evaluation. Its four
predictors total 2.9B parameters, and the complete system runs and streams on a
single consumer GPU.

In speech, the surrounding text and the preceding delivery both carry information about how the next words should sound. We therefore start from a pretrained language model,
which attends to both sources of context, and predict discrete audio tokens at a
low frame rate: at 12.5\,Hz one minute of audio occupies 750 tokens, which lets
the model retain substantial text and audio context. The audio representation is
DualCodec, a residual vector quantizer that biases
its first codebook toward semantics \citep{li2025dualcodec}. That codebook carries content and timing by itself,
while the higher-index codebooks, each encoding residual detail not captured by the preceding codebooks,
add acoustic detail. \modelname{} keeps that codebook and the next three, and
discards the four finest.

\begin{samepage}
Assuming that the first codebook largely determines prosody, we allocate most
model capacity to it. A Qwen3-1.7B-derived transformer predicts it; its output
carries the intonation contour and its length sets the utterance duration.
Three Qwen3-0.6B-derived transformers with
progressively fewer blocks then add the acoustic refinements in order
\citep{qwen3technicalreport}.
\par
\end{samepage}

DualCodec's decoder is not causal, so audio near a chunk boundary depends on
frames that lie beyond it. We therefore map overlapping DualCodec
reconstructions into the VibeVoice acoustic latent space and decode them with
VibeVoice's causal decoder \citep{peng2026vibevoice}, which lets stable audio
be emitted before generation has finished.

We evaluate on audiobook reading. Read prose exposes the failures that matter
for naturalness: monotony, misplaced emphasis, and phrase boundaries that fall
in the wrong place are audible in continuous reading in a way they are not in
isolated words. In a blinded listening study over a corpus of 400 passages,
listeners preferred \modelname{}'s pauses, emphasis, and rhythm over those of
Fish Audio S2 Pro in 70.25\% of comparisons. On an LLM-as-a-judge benchmark over
the same passages, \modelname{} matches ElevenLabs Flash v2.5 on prosody and
outperforms the April 2026 Gradium API and Cartesia Sonic 3. These evaluations
cover English reading; they do not establish
German or broader multilingual quality, voice similarity, or conversational~performance.

The weights are released under the Tontaube license, which permits research
use and qualifying commercial use subject to its revenue and service
restrictions. They are released on Hugging Face alongside an
inference implementation that streams audio as it is generated. The remainder of this report describes
the representation, the four predictors, and their input contract
(Section~\ref{sec:architecture}); the position scheme and the chunking that
keeps context bounded (Section~\ref{sec:positions}); inference, streaming
reconstruction, and serving performance (Section~\ref{sec:generation}); the
evaluations and results (Section~\ref{sec:evaluation}); and the limitations and
release terms (Section~\ref{sec:limits}).

\section{Related Work}
\label{sec:related}

Recent text-to-speech systems increasingly pair a pretrained language-model
backbone with a discrete audio codec and generate codec tokens using the
autoregressive next-token objective employed in language-model pretraining. Fish-Speech
\citep{liao2024fishspeech} and Qwen3-TTS \citep{qwen2026tts} both take this
route, as does \modelname{}. Qwen3-TTS is the closest published system: it
builds on the Qwen3 family and uses a 12.5\,Hz multi-codebook tokenizer whose
first layer carries semantic content and whose later layers carry acoustic
detail.

What differs between such systems is how the stacked codebooks are factorized.
One approach interleaves them into a single output stream with a delay pattern,
producing all codebooks in one autoregressive pass
\citep{copet2023musicgen,lyth2024parler}. Another assigns the coarsest codebook
to a large backbone and the residuals to one small shared module that runs per
frame, as in Fish-Speech's Fast Transformer and Qwen3-TTS's multi-token
prediction module; capacity for the residual codebooks is then small and shared.
\modelname{} instead assigns a separate model to each codebook and sizes them
independently. This design requires four checkpoints and sequential execution
within each chunk; in return, the acoustic stages can be sized independently, carry no state
across chunk boundaries, and be scheduled separately at serving time. Early
experiments showed better convergence when each codebook was assigned a separate
model, which motivated the factorization used here.

Systems that model continuous latents instead, such as VibeVoice
\citep{peng2026vibevoice}, predict the next latent with a diffusion head rather
than the next code from a fixed vocabulary. The codec architecture is also
adopted from prior work: quantizers biased toward linguistic content in their
first layer have appeared in prior systems
\citep{zhang2023speechtokenizer,defossez2024moshi}, and \modelname{} uses
DualCodec \citep{li2025dualcodec} unchanged.

Three aspects of \modelname{} are less common: its character-level text
tokenization, position assignment, and streaming reconstruction. We assign
rotary positions by when a token occurs
rather than by where it sits in the sequence, so corresponding frames of
different streams share a coordinate and text and audio share one logical position axis. Paired
boundary markers keep that timeline aligned across chunks, so a rolling window
can discard completed ones and transformer context stays bounded however long
the passage is. And because DualCodec's decoder is not causal,
streaming is obtained by re-encoding overlapping reconstructions into
VibeVoice's acoustic latent space for causal decoding rather than by training a causal codec as
Qwen3-TTS does.

An earlier model of ours, TontaubeV0 \citep{tontaube2026v0}, was developed
concurrently with Qwen3-TTS and released through an API rather than as weights.
It served as the prototype for the design developed further~in~\modelname{}.

\clearpage
\section{Model Architecture}
\label{sec:architecture}

\modelname{} models speech as a sequence of discrete codes and generates them
autoregressively. Given spoken-form text \(X\), optional reference audio encoded
as prompt streams \(P^{0:3}\), and language/style controls left implicit in the
notation, the task is to sample from \(p(C^{0:3} \mid X, P^{0:3})\), where
\(C^{0:3}\) is the stack of retained codec streams.
Because the streams are
stacked rather than sequential, several factorizations are possible;
\modelname{}'s is described in Section~\ref{sec:factorization}.

\subsection{Speech representation}
\label{sec:representation}

\modelname{} operates on the \texttt{12hz\_v1} configuration of DualCodec
\citep{li2025dualcodec}, which encodes 24\,kHz audio into eight residual
vector-quantized streams at 12.5\,Hz. DualCodec quantizes its first layer from
self-supervised w2v-BERT-2.0 features \citep{chung2021w2vbert}, which carry phonetic and linguistic
information, and the remaining seven layers from acoustic residuals. We call the first stream the semantic stream and the
others the acoustic streams. The semantic codebook holds 16,384 entries and
each acoustic codebook holds 4,096, giving 14 bits per semantic frame and 12
bits per acoustic frame. Decoded on its own, the semantic stream already yields
intelligible speech with recognizable phrasing and intonation, though with
substantially reduced acoustic detail.

\modelname{} retains the semantic stream and the first three acoustic streams,
yielding 625\,bit/s, compared with 1,225\,bit/s for the full stack. The four finest
refinements are omitted because informal listening indicated diminishing
improvements in audio quality from the later streams. Throughout, \(t\) indexes frames on this 12.5\,Hz clock. Each frame
holds one token per retained stream, so a stream of \(L\) tokens spans \(L\)
frames, or \(L/12.5\) seconds.

\subsection{Four-stage codec generation}
\label{sec:factorization}

Let \(X\) be the text, \(P^i\) the reference-audio tokens for stream \(i\), and
\(C^i=(c^i_1,\ldots,c^i_L)\) the tokens generated for that stream. We write
\(C^{0:i}=(C^0,\ldots,C^i)\), \(C^{<i}=(C^0,\ldots,C^{i-1})\), with
\(C^{<0}\) empty, and \(P^{0:i}=(P^0,\ldots,P^i)\). The chain rule gives the
exact coarse-to-fine factorization
\begin{equation}
p(C^{0:3}\mid X,P^{0:3})
=\prod_{i=0}^{3}p(C^i\mid X,P^{0:3},C^{<i}).
\label{eq:chain-factorization}
\end{equation}

\modelname{} restricts the conditioning structure in
Equation~\ref{eq:chain-factorization}: stage \(i\) receives only the prompt
streams \(P^{0:i}\). Its model distribution \(q\) therefore factorizes as
\begin{equation}
q(C^{0:3}\mid X,P^{0:3})
=\prod_{i=0}^{3}q_i(C^i\mid X,P^{0:i},C^{<i}).
\label{eq:model-factorization}
\end{equation}
For \(i<3\), the stage-\(i\) factor is invariant to \(P^{i+1:3}\),
corresponding to the conditional-independence assumption
\(C^i\perp P^{i+1:3}\mid X,P^{0:i},C^{<i}\) under \(q\).

We write \(\cb{i}\) for the model \(q_i\) and refer to the four models as
stages. The other design choices are to model each factor with its own network
and to generate the factors in order without allowing a later stage to revise
an earlier stream.
\(\cb{0}\) reads the text and its reference-prompt stream and produces the
semantic stream, stopping when it emits a boundary token; the number of tokens
it produced is \(L\). Each of \(\cb{1}\), \(\cb{2}\), and \(\cb{3}\) then
produces exactly \(L\) tokens for its own stream, conditioned on the text, its
available prompt streams, and every stream already completed below it. The four
streams therefore share one timeline, fixed once by \(\cb{0}\) and unchanged
afterwards.

The ordering is an inductive bias rather than a restriction on the exact chain
rule. No stage sees tokens from a higher stream.
Once a candidate semantic stream \(C^0\) is accepted, its tokens and timeline
remain fixed. The acoustic stages add residual acoustic detail and may
compensate for lower-stage quantization errors, but they do not resample
\(C^0\) or change its length. The first stage therefore strongly constrains
pronunciation, phrasing, and duration. By default, the later stages run at temperature
zero, so their decoded outputs are deterministic given \(X\), \(P^{0:3}\), and
\(C^0\).

\subsection{Input and output}
\label{sec:io}

Each stage consumes one flat token sequence and emits tokens from a single
codebook. The stages share a common alphabet of character IDs for spoken text,
inherited BPE pieces for the language and style labels, and dedicated IDs for
structure (\texttt{<|text\_split|>}, \texttt{<|audio\_split|>}, PAD, and the
row terminators). Each stage augments the common alphabet with the codec
vocabularies of every stream up to and including the one it generates;
consequently, the input alphabet of each stage after \(\cb{0}\) contains that of
the preceding stage. Each
stage emits tokens from its own codebook, plus the structural tokens it is
allowed to produce. For \(\cb{0}\) that is 16,384 semantic tokens with a row
terminator and \texttt{<|audio\_split|>}; for \(\cb{1}\) to \(\cb{3}\) it is
4,096 acoustic tokens and nothing else. Logits outside the target vocabulary
are masked at sampling time.

Table~\ref{tab:special-tokens} lists the structural tokens.
\texttt{<|im\_start|>}, \texttt{<|im\_end|>}, and PAD
(\texttt{<|endoftext|>}) are inherited from Qwen;
\texttt{<|end\_of\_speech|>}, the split markers, and the codec IDs were added.

\begin{table}[H]
\centering
\caption{Structural tokens. The acoustic stages are length-matched to \(C^0\)
and emit no structural tokens of their own.}
\label{tab:special-tokens}
\vspace{8pt}
\small
\begin{tabular}{llll}
\toprule
Token & Read by & Emitted by & Purpose \\
\midrule
\texttt{<|im\_start|>}, \texttt{<|im\_end|>} & \(\cb{0}\)--\(\cb{3}\) & --- & delimit the control block \\
\texttt{<|text\_split|>} & \(\cb{0}\)--\(\cb{3}\) & --- & aligned text boundary \\
\texttt{<|audio\_split|>} & \(\cb{0}\)--\(\cb{3}\) & \(\cb{0}\) & corresponding boundary in \(C^0\) \\
\texttt{<|end\_of\_speech|>} & \(\cb{1}\)--\(\cb{3}\) & \(\cb{0}\) & stops \(\cb{0}\); terminates its completed row \\
\(\mathrm{PAD}\) & \(\cb{0}\)--\(\cb{3}\) & --- & opens an audio row \\
\bottomrule
\end{tabular}
\end{table}

For \(\cb{0}\), a minimal request without reference audio is serialized as
follows, with \sep{} marking a token boundary and \textvisiblespace{} a literal
space character. The control block is BPE-encoded; within the spoken-text
segment, each character occupies one token.

\begin{center}
\small\ttfamily
\begin{tabular}{@{}ll@{}}
in & <|im\_start|>\sep english\sep \textvisiblespace{}:\sep \textvisiblespace{}audi\sep obook\sep <|im\_end|>\sep \textbackslash n\sep H\sep i\sep \textvisiblespace\sep t\sep h\sep e\sep r\sep e\sep .\sep \textbackslash n\sep PAD \\
out & \(c^0_1\)\sep \(c^0_2\)\sep \(\ldots\)\sep \(c^0_L\)\sep <|end\_of\_speech|> \\
\end{tabular}
\end{center}

\noindent
The acoustic stages reuse that prefix and append one line per completed row.
The full layout for \(\cb{3}\), with the prompt block omitted, is:

\begin{center}
\small
\begin{tabular}{@{}l@{\quad}l@{}}
in & \ttfamily <|im\_start|>\sep english\sep \textvisiblespace{}:\sep \textvisiblespace{}audi\sep obook\sep <|im\_end|>\sep \textbackslash n \\
& \ttfamily H\sep i\sep \textvisiblespace\sep t\sep h\sep e\sep r\sep e\sep .\sep \textbackslash n \\
& \ttfamily PAD\sep \(c^0_1\)\sep \(\ldots\)\sep \(c^0_L\)\sep <|end\_of\_speech|> \\
& \ttfamily PAD\sep \(c^1_1\)\sep \(\ldots\)\sep \(c^1_L\) \\
& \ttfamily PAD\sep \(c^2_1\)\sep \(\ldots\)\sep \(c^2_L\) \\
& \ttfamily PAD \\
out & \ttfamily \(c^3_1\)\sep \(\ldots\)\sep \(c^3_L\) \\
\end{tabular}
\end{center}

\noindent
\(\cb{1}\) and \(\cb{2}\) are the same with fewer rows: \(\cb{1}\) has only
the \(C^0\) row before its target, and \(\cb{2}\) has \(C^0\) and
\(C^1\).

\noindent
Only \(C^0\) is followed by \texttt{<|end\_of\_speech|>}, since it is the one
row whose length is not known in advance; \(C^1\) and \(C^2\) are
length-matched to it. Appendix~\ref{app:chunked-example} shows how text and
audio split markers delimit a continuation. In ordinary chunk-local acoustic
refinement, \(\cb{1}\)--\(\cb{3}\) encounter neither split marker. When a
continuation prefix is supplied, they receive \texttt{<|text\_split|>} between
the prefix and current text and \texttt{<|audio\_split|>} at the corresponding
boundary in \(C^0\).

The text to be spoken \(X=(x_1,\ldots,x_n)\) is tokenized one character at a
time. Rather than introduce a character vocabulary, \modelname{} reuses IDs
the inherited tokenizer already produces: each character is encoded on its own, and
only the first resulting ID is kept, even when the tokenizer returns several.
Writing \(B\) for the inherited Qwen
tokenizer and \(\operatorname{first}\) for the first element of a token
sequence,
\begin{equation}
\tau(X)=\bigl(\operatorname{first}(B(x_1)),\ldots,
                 \operatorname{first}(B(x_n))\bigr),
\label{eq:char-tokenization}
\end{equation}
so ordinary content occupies exactly \(n\) tokens and no merge crosses a
character boundary. Two things follow. Chunk sizes, lookahead windows, and text
positions are measured in a unit that does not depend on neighbouring words.
And pronunciation is learned over a few hundred character IDs rather
than tens of thousands of sparsely observed subword types, so character-to-sound
patterns are reused across words and spellings and inference text is composed
almost entirely of well-observed IDs.
Case is preserved; the released inference path does not lowercase input text.
Its mechanical sanitizer strips surrounding whitespace, collapses runs of
spaces and tabs to one space, preserves a single newline and at most one blank
line, and maps other vertical whitespace to newlines. It appends a period to
each nonempty string it sanitizes unless the string's final character is
\texttt{.}, \texttt{!}, \texttt{?}, \texttt{;}, \texttt{:}, or \texttt{,}.
When the inference engine builds a model input, it strips each resulting chunk
again, adds exactly one leading space to every chunk but the first, and adds a
newline only to the last; callers do not supply these~boundary~cues.

Ordinary spoken text is character-tokenized; subword tokenization is confined
to control syntax. The main control block has the form
\texttt{<|im\_start|>}\allowbreak\texttt{language : style}\allowbreak
\texttt{<|im\_end|>} followed by a newline, with the label text encoded by the
inherited tokenizer and the delimiters as dedicated IDs. The released interface
supports the styles \texttt{audiobook}, \texttt{conversational}, and
\texttt{agentic}; the supported language labels are listed in
Section~\ref{sec:limits}. Labels outside the released sets are not supported.

Reference audio is optional and requires no transcript. Up to roughly 60
seconds are encoded by DualCodec using four quantizers, giving aligned prompt
streams of equal length. Stage \(i\) is conditioned on the prompt streams
\(P^0,\ldots,P^i\); in particular, \(\cb{0}\) receives only the semantic
prompt. Before serialization, these streams are jointly truncated to a
stage-specific maximum of 750, 300, 150, or 100 frames for \(\cb{0}\) through
\(\cb{3}\), respectively. This corresponds to 60, 24, 12, or 8 seconds at
12.5\,Hz. They are then placed back-to-back before the control block, without
separator or PAD tokens. Disjoint codec-token ranges preserve stream identity,
while logical positions restart at 1 for each stream. PAD tokens are instead
used to open the post-text audio rows. A request without reference audio omits
the prompt block entirely.

\paragraph{Text verbalization.}
\modelname{} expects \(X\) already in spoken form. Digits, dates, currencies,
and abbreviations are not reliably pronounced from their written shape, so text
containing them must be verbalized first: \emph{1984} must be supplied as
\emph{nineteen eighty-four} or \emph{one thousand nine hundred eighty-four},
depending on the intended meaning, and the model cannot reliably make that choice from the
characters alone. A separately released English verbalizer is available
for this purpose. Keeping it separate means its output can be inspected and
corrected before synthesis, and callers who need exact control can supply
spoken-form text directly; Appendix~\ref{app:verbalizer} gives the details. Text in the
other supported languages must be supplied already verbalized.

\subsection{Predictor architecture}
\label{sec:predictors}

\begin{table}[t]
\centering
\caption{Released predictor architecture and parameter accounting. Stored
parameter counts are the total number of tensor elements in the released
safetensors checkpoints. Restricted-input counts subtract input
embedding rows unused by the documented input grammar; all
output-head rows remain active. DualCodec, VibeVoice, and the optional
verbalizer are excluded.}
\label{tab:parameters}
\vspace{8pt}
\small
\setlength{\tabcolsep}{4.2pt}
\resizebox{0.99\textwidth}{!}{%
\begin{tabular}{lrrrrrrr}
\toprule
Model & Blocks & Width & FFN & Embedding rows & Retained input rows &
Stored parameters & Restricted-input parameters \\
\midrule
\(\cb{0}\) & 28 & 2,048 & 6,144 & 168,057 & 34,955 &
1,829,116,930 & 1,556,524,034 \\
\(\cb{1}\) & 16 & 1,024 & 3,072 & 172,153 & 39,052 &
448,960,512 & 312,665,088 \\
\(\cb{2}\) & 8 & 1,024 & 3,072 & 176,249 & 43,148 &
327,307,264 & 191,011,840 \\
\(\cb{3}\) & 4 & 1,024 & 3,072 & 180,345 & 47,244 &
268,577,792 & 132,282,368 \\
\midrule
Total & 56 & --- & --- & --- & --- &
2,873,962,498 & 2,192,483,330 \\
\bottomrule
\end{tabular}
}
\end{table}

\(\cb{0}\) uses a Qwen3-1.7B-derived backbone; \(\cb{1}\) to \(\cb{3}\) use
Qwen3-0.6B-derived backbones with progressively fewer blocks
(Table~\ref{tab:parameters}) \citep{qwen3technicalreport}. The inherited input
embedding table is extended to support the stage-specific vocabulary described
in Section~\ref{sec:io} and optimized jointly with the transformer backbone.

The inherited language-model head is discarded. In its place each stage uses a
newly initialized, untied, two-layer audio head: it expands the transformer
state to width 4,096, applies a Mish nonlinearity, and projects to that stage's
audio vocabulary.

\paragraph{Parameter accounting.}
Table~\ref{tab:parameters} reports two parameter counts per stage. The stored count sums
every element in the released checkpoint. The checkpoints keep Qwen's full
embedding table, but the input scheme of Section~\ref{sec:io} can only produce
a fraction of those IDs; the restricted count subtracts the rows that are never
reachable, removing 681,479,168 parameters across the four models, or roughly 1.4\,GB at
bf16. The rows are kept in the release so that later tuning can widen the
control-tag vocabulary; implementations that enforce the documented scheme
can omit the unreachable rows from the embedding table.

\subsection{Training}
\label{sec:training}

All four predictors were trained exclusively with supervised fine-tuning (SFT)
on approximately 200,000 hours of paired speech and text across seven
languages, predominantly from public-domain audiobook recordings and openly
released speech corpora. We do not disclose further details of the data
composition, training schedule, or hyperparameters in this report.

\section{Positions and Long-Form Layout}
\label{sec:positions}

Positions are assigned to match an intuitive
notion of when each token occurs rather than where it was serialized, which
builds temporal alignment into the encoding instead of leaving it to be learned.
Passages longer than one context window are generated in chunks under a
scheme that keeps the transformer's context bounded however long the passage
grows.

\subsection{Positions}

The inputs of Section~\ref{sec:io} are a set of rows that share a time axis:
the prompt streams, the text, and every stream already completed for the
current chunk. We use \emph{row} for any one of these sequences, whether text
or codec. A decoder-only transformer consumes one flat sequence.
\modelname{} therefore separates physical order from logical time. The tenth
semantic frame and the tenth acoustic frame describe the same moment of speech
but can be hundreds of token indices apart once serialized; giving them the same
coordinate sets their relative positional offset under RoPE to zero, whatever the
layout, since it depends only on coordinate differences \citep{su2024rope}.
Text shares the timeline, so character positions remain close to those of the
corresponding audio frames; the
two clocks run at different rates and are realigned at each chunk boundary
(Section~\ref{sec:markers}).

For position assignment, the control block and spoken text are treated as a
single text row. Writing \(\ell_P\) for the length of one prompt stream, \(N\) for
the total length of this row, and \(L\) for the audio length, the coordinates
are

\begin{center}
\small
\setlength{\extrarowheight}{2pt}
\begin{tabular}{@{}ll@{}}
\toprule
Block & Positions \\
\midrule
each prompt stream \(P^j\) & \(1,\ldots,\ell_P\) \\
each PAD opening an audio row & \(\ell_P\) \\
text row & \(\ell_P+1,\ldots,\ell_P+N\) \\
each codec row \(C^j\) & \(\ell_P+1,\ldots,\ell_P+L\) \\
\texttt{<|end\_of\_speech|>} after \(C^0\) & \(\ell_P+L+1\) \\
\bottomrule
\end{tabular}
\end{center}

\noindent
Three things follow. The prompt streams overlay one another rather than running
end to end, so four streams of \(\ell_P\) frames occupy \(\ell_P\) coordinates, not
\(4\ell_P\). Under this convention, text and audio both start at \(\ell_P+1\), thereby
placing them on the same logical timeline. And frame \(t\) carries position
\(\ell_P+t\) in every row,
whether it comes from a completed lower stream or the one being generated, so
physical order determines causal visibility, while shared coordinates encode
simultaneity. These are within-chunk coordinates; the chunk
markers sit at boundaries given by Equation~\ref{eq:markers} below.
Figure~\ref{fig:positions} plots serialized sequence index \(r\) against
logical position \(\rho\), for \(\cb{0}\) mid-passage and for~the~full~\(\cb{3}\)~prefix.

\begin{figure}[t]
\centering
\resizebox{0.985\textwidth}{!}{%
\begin{tikzpicture}[
  x=1cm,y=.68cm,
  seg/.style={draw,rounded corners=2.5pt,minimum height=.72cm,
              font=\scriptsize,align=center,line width=.65pt},
  prompt/.style={seg,fill=PromptFill,draw=PromptStroke},
  txt/.style={seg,fill=TextFill,draw=TextStroke},
  marker/.style={seg,fill=MarkerFill,draw=MarkerStroke},
  audio/.style={seg,fill=AudioFill,draw=AudioStroke},
  target/.style={seg,fill=TargetFill,draw=TargetStroke},
  pad/.style={seg,fill=PadFill,draw=PadStroke},
  axis/.style={-{Latex[length=2.1mm]},draw=black!55,line width=.7pt},
  jump/.style={-{Latex[length=2mm]},draw=DiagramPath,line width=.9pt,dashed},
  guide/.style={draw=black!13,line width=.5pt},
  note/.style={font=\scriptsize,text=black!65,align=left}
]
% CB0 semantic generation.
\node[font=\small\bfseries,anchor=west] at (1.55,7.18)
  {(a) \cb{0}: semantic generation for an internal chunk};

\draw[axis] (1.25,1.65) -- (24.8,1.65)
  node[below,font=\scriptsize] {serialized sequence index $r$};
\draw[axis] (1.25,1.65) -- (1.25,6.85);
\node[font=\scriptsize,rotate=90] at (-.35,4.25)
  {logical RoPE position $\rho$};
\foreach \y/\lab in {2.05/$1$,2.65/$\ell_P$,3.35/$S_0$,4.05/$M_0$,
                       4.65/$M_0{+}1$,5.85/$M_1$,6.35/$M_1{+}1$} {
  \draw[guide] (1.25,\y) -- (24.35,\y);
  \node[font=\scriptsize,anchor=east] at (1.08,\y) {\lab};
}

\node[prompt,minimum width=1.75cm] (f01) at (2.325,2.05)
  {$P^0$ prompt\\range $1{:}\ell_P$};
\node[txt,minimum width=3.25cm] (f02) at (5.075,3.35) {};
\node[font=\scriptsize] at ([xshift=-.90cm]f02.center) {\strut system tag};
\node[font=\scriptsize] at ([xshift=.625cm]f02.center) {\strut previous text};
\draw[draw=TextStroke,densely dotted,line width=.6pt]
  ([xshift=-.225cm]f02.north) -- ([xshift=-.225cm]f02.south);
\node[marker,minimum width=.78cm] (f04) at (7.34,4.05) {$T_0$};
\node[txt,minimum width=2.25cm] (f05) at (9.105,4.65) {current text};
\node[marker,minimum width=.78cm] (f06) at (10.87,5.85) {$T_1$};
\node[txt,minimum width=2.20cm] (f07) at (12.61,6.35) {short future text};
\node[pad,minimum width=.70cm] (f08) at (14.31,2.65) {PAD};
\node[audio,minimum width=2.05cm] (f09) at (15.935,3.35) {previous $C^0$};
\node[marker,minimum width=.78cm] (f10) at (17.60,4.05) {$A_0$};
\node[target,minimum width=2.55cm] (f11) at (19.515,4.65)
  {generate current $C^0$};
\node[marker,minimum width=1.18cm] (f12) at (21.63,5.85)
  {$A_1$};

\foreach \a/\b in {f01/f02,f02/f04,f04/f05,f05/f06,
                     f06/f07,f07/f08,f08/f09,f09/f10,f10/f11}
  \draw[jump] (\a.east) -- (\b.west);
\draw[jump] (f11.east) -- (f12.west);

% CB3 acoustic refinement: full repeated-row pattern.
\node[font=\small\bfseries,anchor=west] at (1.55,.72)
  {(b) \cb{3}: every completed lower codebook adds one aligned row};

\begin{scope}[yshift=-4.3cm]
\draw[axis] (1.25,1.65) -- (24.8,1.65)
  node[below,font=\scriptsize] {serialized sequence index $r$};
\draw[axis] (1.25,1.65) -- (1.25,5.45);
\node[font=\scriptsize,rotate=90] at (-.35,3.55)
  {logical RoPE position $\rho$};
\foreach \y/\lab in {2.05/$1$,3.10/$\ell_P$,4.15/$\ell_P{+}1$,
                       5.15/$\ell_P{+}L{+}1$} {
  \draw[guide] (1.25,\y) -- (24.35,\y);
  \node[font=\scriptsize,anchor=east] at (1.08,\y) {\lab};
}
\node[prompt,minimum width=1.30cm] (g1) at (2.10,2.05) {$P^0$};
\node[prompt,minimum width=1.30cm] (g2) at (3.62,2.05) {$P^1$};
\node[prompt,minimum width=1.30cm] (g3) at (5.14,2.05) {$P^2$};
\node[prompt,minimum width=1.30cm] (g4) at (6.66,2.05) {$P^3$};
\node[txt,minimum width=3.27cm] (g5) at (9.165,4.15) {};
\node[font=\scriptsize] at ([xshift=-.83cm]g5.center) {\strut system tag};
\node[font=\scriptsize] at ([xshift=.70cm]g5.center) {\strut current text};
\draw[draw=TextStroke,densely dotted,line width=.6pt]
  ([xshift=-.135cm]g5.north) -- ([xshift=-.135cm]g5.south);
\node[pad,minimum width=.65cm] (g7) at (11.345,3.10) {PAD};
\node[audio,minimum width=1.55cm] (g8) at (12.665,4.15) {complete $C^0$};
\node[marker,minimum width=.82cm] (g9) at (14.15,5.15)
  {$\mathrm{EOS}$};
\node[pad,minimum width=.65cm] (g10) at (15.30,3.10) {PAD};
\node[audio,minimum width=1.55cm] (g11) at (16.62,4.15) {complete $C^1$};
\node[pad,minimum width=.65cm] (g12) at (17.94,3.10) {PAD};
\node[audio,minimum width=1.55cm] (g13) at (19.26,4.15) {complete $C^2$};
\node[pad,minimum width=.65cm] (g14) at (20.58,3.10) {PAD};
\node[target,minimum width=2.10cm] (g15) at (22.175,4.15)
  {generate $C^3$};
\foreach \a/\b in {g1/g2,g2/g3,g3/g4,g4/g5,g5/g7,g7/g8,
                     g8/g9,g9/g10,g10/g11,g11/g12,g12/g13,g13/g14,g14/g15}
  \draw[jump] (\a.east) -- (\b.west);
\end{scope}

\end{tikzpicture}
}
\vspace{1mm}
{\scriptsize
\colorbox{PromptFill}{\strut voice prompt}\quad
\colorbox{TextFill}{\strut text/control}\quad
\colorbox{MarkerFill}{\strut boundary marker}\quad
\colorbox{AudioFill}{\strut known audio}\quad
\colorbox{TargetFill}{\strut generated audio}\quad
\colorbox{PadFill}{\strut PAD}}
\caption{Serialization index \(r\) versus logical RoPE position \(\rho\). Boxes
mark block starts; within each block, \(r\) and \(\rho\) advance by one per
token. Dashed arrows trace physical serialization; their endpoints expose
forward jumps, resets, and reused logical coordinates. Panel (a) ends an
internal chunk with \(A_1\) at \(M_1\). Panel (b) shows the full \(\cb{3}\)
layout; \(\cb{1}\) and \(\cb{2}\) use shorter prefixes. A dotted divider changes
labels, not positions. Here, \(S_k\) and \(M_k\) are the start and boundary
coordinates of chunk \(k\), while \(T_k\) and \(A_k\) are its paired text and
audio markers. In panel (b), \(\mathrm{EOS}\) denotes the
\texttt{<|end\_of\_speech|>} token. It terminates the completed \(C^0\) row and
does not necessarily mark the end of the passage~being~refined.}
\label{fig:positions}
\end{figure}

\subsection{Chunk boundaries}
\label{sec:markers}

Long passages are split into paired text and audio segments. Text advances one
position per character and audio one per frame, so within a segment the two
drift apart. Each boundary \(k\) is therefore given a single position
\(M_k\) that both jump to, marked by \texttt{<|text\_split|>} in the text and
\texttt{<|audio\_split|>} in the audio. Boundaries advance monotonically, and because character rate and frame rate are
close, the positions grow roughly in proportion to elapsed speech.

\begin{samepage}
For segment \(k\) starting at logical position \(S_k\), with text length
\(n^x_k\), available semantic-stream length \(n^a_k\), and a fixed offset
\(\delta\), \(M_k\) and the next start are
\begin{equation}
M_k=S_k+\max(n^x_k+\delta,n^a_k),\qquad S_{k+1}=M_k+1.
\label{eq:markers}
\end{equation}
Thus, the larger of \(n^x_k+\delta\) and \(n^a_k\) determines the boundary
position. The offset \(\delta\)
reserves headroom before the semantic length is known, since text and audio
lengths can differ in either direction; the released checkpoints use
\(\delta=25\) character positions, a unit that stays stable across passages
because text is tokenized per character (Section~\ref{sec:io}).
\end{samepage}

\subsection{Bounded context}

\modelname{} treats prosodic context as predominantly local, so \(\cb{0}\)
keeps one preceding chunk rather than the whole passage. The retained chunk
provides local context intended to support continuity across the boundary,
while the transformer's input context remains bounded as the passage grows. For an internal
chunk, \(\cb{0}\)'s text window holds the
previous chunk, the current chunk, and a short lookahead into the next; its
audio context holds only the previous chunk's semantic tokens \(C^0\).
Appendix~\ref{app:chunked-example} shows the layout. Once a chunk completes,
the oldest text--audio chunk pair is discarded from the transformer's input
window and the window moves on. The paired markers keep the timeline from drifting each time the window
moves.

The leading-space and final-newline cues tell \(\cb{0}\) where the window lies
in the passage: the space distinguishes noninitial chunks, and the newline
distinguishes the final chunk from an internal one. The released splitter caps
chunks at 350 characters. On each overlong remainder it cuts after the last
period, exclamation mark, question mark, semicolon, colon, or comma within the
cap; if none occurs, it cuts at the last space, tab, or newline, and if no such
whitespace occurs it splits hard at 350 characters. These punctuation marks
form one priority class rather than separate sentence and comma passes. The
ceiling is an inference-side context margin, not the training chunk rule:
training grouped forced-alignment segments under a joint text-and-audio
sequence budget.

\subsection{Serving}

The inference repository ships vLLM adapters for the four predictors. The layout
above assigns positions that do not follow the physical sequence index, whereas
vLLM's decode path assumes they do. The adapters resolve this by adding one
uniform constant to every prefill coordinate, chosen so that the next
identity-indexed decode token has the intended relative positions with respect
to the prefill tokens, after accounting for any skipped marker position. The
absolute coordinates then differ from those
seen in training, but rotary attention depends only on coordinate differences
\citep{su2024rope}, so every relative offset, and therefore the learned
alignment, is preserved. The custom layout can then use vLLM's standard decode
and KV-cache path with no change to the weights
\citep{kwon2023pagedattention,vllm2026registration}.

\section{Inference}
\label{sec:generation}

Generation runs coarse to fine, as Section~\ref{sec:factorization} describes.
\(\cb{0}\) stops on \texttt{<|audio\_split|>}, on
\texttt{<|end\_of\_speech|>}, or on a configured generation limit. The sampled
stopping symbol is removed from the returned codec codes; the completed
\(C^0\) conditioning row retains the corresponding boundary marker. Each
acoustic stage then generates exactly \(L\) tokens from the completed rows
below it.

\subsection{Chunk-local acoustic refinement}

\(\cb{1}\) to \(\cb{3}\) carry no autoregressive state across chunk
boundaries. Each conditions only on its prompt streams, the current chunk's
text and complete \(C^0\), and the lower acoustic rows already finished for
that same chunk. Two things follow. Voice drift cannot accumulate through
cross-chunk acoustic state, and once the semantic stream exists for several
chunks, those chunks can be refined concurrently in separate calls or together
in one batch. Within a chunk the stages remain strictly ordered: \(C^1\) before
\(C^2\), and \(C^2\) before \(C^3\).

\subsection{Reconstruction}
\label{sec:reconstruction}

The four retained streams are decoded by DualCodec to 24\,kHz audio. That
decoder is not causal, so waveform samples near a cut depend on codec frames
beyond it, and independently decoding chunks and joining the waveforms could
introduce an audible seam at each boundary \citep{dualcodecsoftware}. \modelname{}
therefore joins the reconstructed windows in a latent space instead: overlapping DualCodec
reconstructions are re-encoded by the VibeVoice acoustic tokenizer, the
context-padded central frames of each window are retained, and a single causal
VibeVoice decoder cache is carried across the joined sequence
\citep{microsoftvibevoiceacoustic}. Only VibeVoice's acoustic encoder and
decoder are used, not its language model or diffusion head.

For a completed response, reconstruction uses 30-second content windows padded
with six seconds of DualCodec context on each available side. VibeVoice encodes
the padded waveform, but only the latent frames inside the central window are
kept. Those regions are concatenated and decoded in groups with a single
bounded-size causal VibeVoice convolutional decoder cache carried across the
utterance, so the join occurs in latent space rather than between independently
decoded waveform chunks.

More generally, this construction enables streaming with a noncausal codec
without retraining it: decode overlapping windows, keep only the stable
interiors, and pass them to a causal decoder in its own latent space. This
requires a second codec in the inference path and introduces latency by
withholding a few audio frames.

\subsection{Streaming}

Streaming applies the same construction incrementally. By default, the system
first generates a 40-frame semantic prefix. It withholds five DualCodec frames
at each unstable boundary, leaving 35 frames (2.8 seconds) initially eligible
for emission. As more frames arrive,
the accumulated prefix is decoded and re-encoded, thereby revising the unstable
latent boundary region. Whenever another semantic segment becomes available, it
commits the newly stable, previously unemitted VibeVoice frames at boundaries
aligned to two-second intervals. A persistent, bounded-size convolutional cache
carries the finite causal history needed to continue from the committed
sequence, and the remaining VibeVoice frames are committed when generation
finishes. The system therefore begins emitting audio before generation of the
utterance is complete, without introducing a separately decoded waveform
boundary at any generation boundary.

\subsection{Serving performance}

The measurements below use one NVIDIA GeForce RTX 5090 with weights resident
and the process warmed; startup and model loading are excluded. For a single
input text, the time to first encoded audio is approximately 200\,ms and the
end-to-end real-time factor is 0.08, or about 12.5 times faster
than playback. With eight texts generated concurrently, aggregate throughput
reaches a real-time factor of approximately 0.02, about 50 times real time.

\FloatBarrier

\section{Evaluation}
\label{sec:evaluation}

We evaluate the system's prosody through a blinded listening study on English
reading passages and assess intelligibility separately using automatic speech
recognition (ASR). An LLM-as-a-judge benchmark on the same passages extends the
prosody comparison to further systems. We also test whether generating longer
passages reduces intelligibility or speaker similarity in the final chunk.

\subsection{Blinded listening study}
\label{sec:listening}

\paragraph{Protocol.}
We use Fish Audio S2 Pro as an open-weight reference system with expressive
speech capabilities \citep{liao2026fishaudios2}. Forty listeners recruited
through Prolific each judged ten pairs of \modelname{} and S2 Pro
\citep{fishaudio2026s2pro} recordings using the same voice-cloning reference.
S2 Pro ran locally from the released weights at temperature 0.8 and top-p 0.8,
the defaults of its inference server.
Eligibility required current US residence and English as the primary language.
Forty disjoint sets of ten passages covered each of 400 English book passages
of 250--500 characters exactly once. The passages are sampled from the PG-19
test split, which is derived from Project Gutenberg
\citep{rae2020compressive}. Each listener received one set, with each TTS
system assigned to A in five pairs and B in five. System identities were
hidden. Both systems' recordings were normalized to \(-20\)\,dBFS average level
and presented as 64\,kbps mono MP3s. \modelname{} used semantic temperature
0.65 with text verbalization into spoken form (Appendix~\ref{app:verbalizer});
its recordings were decoded as in Section~\ref{sec:reconstruction}, quiet
pauses were shortened to at most 2.5\,s, and MossFormer2 super-resolution
\citep{zhao2025clearervoice} raised the sample rate to 48\,kHz.

For each pair, listeners answered: ``Which recording do you prefer,
considering how well its pauses, emphasis, and rhythm fit the text?''
They were instructed to disregard voice preference and recording quality.
Both recordings had to play to the end before a choice could be submitted;
replays were permitted. Each pair received one forced-choice judgment.

\begin{figure}[!t]
\centering
\includegraphics[width=0.55\textwidth]{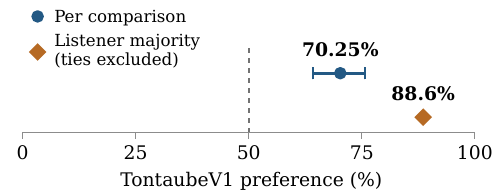}
\caption{Human prosody preference for \modelname{} over Fish Audio S2 Pro:
281/400 pairwise choices and majority preference among 31/35 listeners,
excluding five evenly split listeners. The error bar shows the 95\%
participant-clustered GEE confidence interval for per-comparison preference.}
\label{fig:human-preference}
\end{figure}

\paragraph{Results.}
Listeners preferred \modelname{} in 281 of 400 pairs (70.25\%; 95\% CI:
64.2--75.7\%; Figure~\ref{fig:human-preference}). The interval uses a logistic
generalized estimating equation (GEE) with exchangeable within-listener
correlation and robust standard errors. At the listener level, 31 preferred
\modelname{} in a majority of pairs, four preferred Fish Audio S2 Pro, and five
were evenly split (two-sided exact sign test on the 35 non-tied listeners:
\(p<10^{-5}\)).

Each passage was rated once, so listener and passage effects cannot be
separated. Clip durations differed (median 23.5\,s for \modelname{} versus
20.9\,s for S2 Pro), as did sample rates (48 versus 44.1\,kHz). The study
measures prosody preference, not intelligibility, voice similarity, or
fidelity.

\subsection{LLM-as-a-judge benchmark}
\label{sec:llm-judge}

\paragraph{Protocol.}
The benchmark uses the same 400 passages as the listening study.
\modelname{} renders them in its \texttt{audiobook} mode, at semantic sampling
temperature 0.55 and acoustic temperature zero in every reported comparison;
these renders differ from the recordings used in the listening study. For each
passage, \modelname{} and the comparator synthesize the same reference text,
and each waveform is independently normalized to an average level of
\(-20\)\,dBFS before judging.

For this benchmark, we use Gemini as an order-balanced pairwise judge.
Gemini 3.1 Pro Preview
\citep{google2026gemini31pro} receives the reference text and both waveforms
and judges two dimensions independently: prosody (rhythm, intonation, emphasis,
pacing, and naturalness) and word-by-word text correctness. The prompt
explicitly directs it to ignore voice identity and timbre as well as recording
artifacts, codec artifacts, and overall sound quality; the full instructions are
in Appendix~\ref{app:judge-prompt}. Each pair is judged exactly twice, once in
each presentation order, and \texttt{FIRST}/\texttt{SECOND} outputs are mapped
back to system identity before scoring. A \modelname{} preference, tie, or
comparator preference receives a score of 1, \(1/2\), or 0, respectively, and
the reported preference score is the mean of these values over the 800
judgments. The two calls for a passage are retained separately in the mean
rather than collapsed to one categorical verdict. For uncertainty estimation the
400 passages are bootstrap-resampled as paired clusters, so the 800
order-swapped calls are not treated as independent observations.

This LLM-as-a-judge protocol follows the audio-language-model-as-judge
approach of EmergentTTS-Eval, which reports a Spearman correlation of 0.905
between aggregate human and model-judge system rankings in its study
\citep{manku2025emergentttseval}. We adapt it to order-balanced pairwise
comparisons on a fixed reading set, which gives a repeatable and scalable
alternative to commissioning a listener panel for each comparison.

The Fish Audio comparison uses the same frozen cloning reference for both
systems. In the remaining comparisons \modelname{} uses that reference while
the comparators use fixed provider voices: ElevenLabs \texttt{James}, Gradium
\texttt{QETTJoT4n\_WmpL3w}, and Cartesia
\texttt{79f8b5fb-2cc8-479a-80df-29f7a7cf1a3e}. Voice identity and timbre are
excluded from the judging rubric, but prosody is not fully separable from a
cloned reference. The comparisons with ElevenLabs, Gradium, and Cartesia may
therefore favor \modelname{}, which can inherit aspects of the reference
recording's reading style while those systems use fixed provider voices.

\begin{figure}[!t]
\centering
\includegraphics[width=0.98\textwidth]{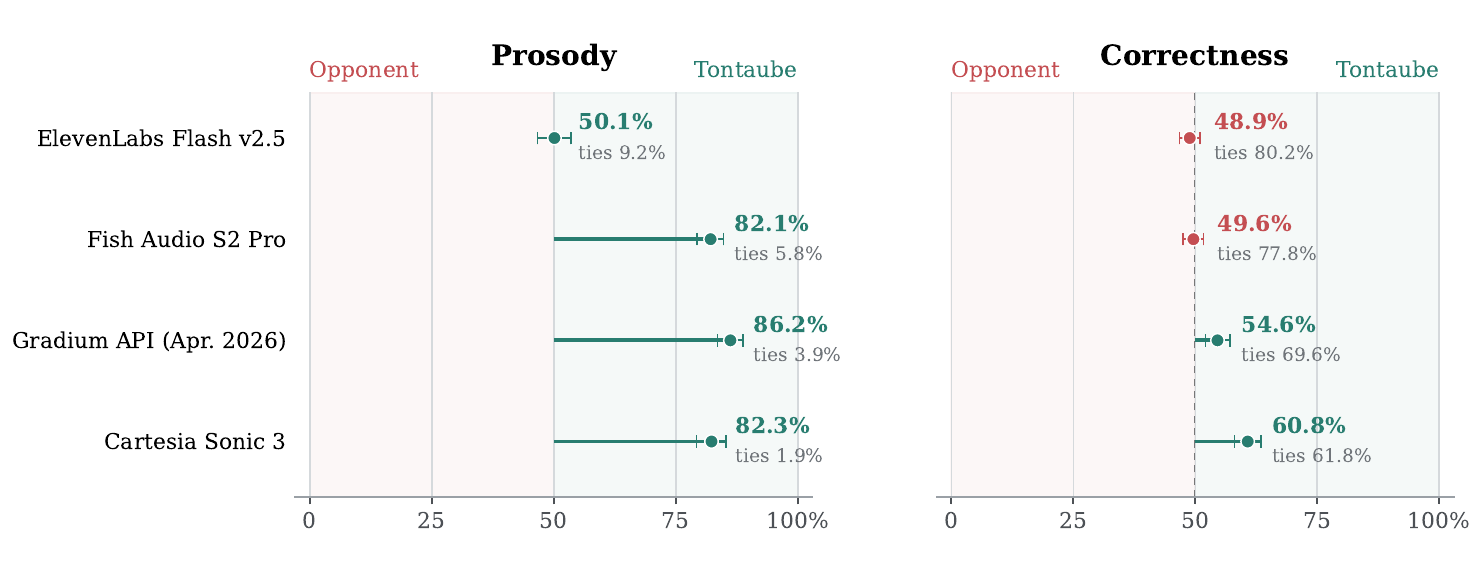}
\caption{LLM-as-a-judge pairwise audiobook-reading benchmark. Points are
TontaubeV1 preference scores over 800 order-balanced judgments; whiskers show
95\% bootstrap intervals obtained by resampling the 400 passages as paired
clusters. A score of 50\% denotes parity. Labels below the points give the tie
rate across individual judge calls.}
\label{fig:pairwise-evaluation}
\end{figure}

\paragraph{Results.}
Against ElevenLabs Flash v2.5 \citep{elevenlabs2026flash25}, the prosody score
is 50.1\% (Figure~\ref{fig:pairwise-evaluation}), with an interval that includes parity; correctness is likewise
statistically indistinguishable from parity at 48.9\%. The prosody scores
against Fish Audio S2 Pro \citep{fishaudio2026s2pro}, the April 2026 Gradium
API \citep{gradium2026tts}, and Cartesia Sonic 3
\citep{cartesia2026sonic3} are 82.1\%, 86.2\%, and 82.3\% respectively. The
correctness interval includes parity against Fish Audio, while the scores
against Gradium and Cartesia favor \modelname{}. Within this English
audiobook-reading benchmark, prosody is therefore comparable to ElevenLabs
Flash v2.5 and ahead of the other three systems.

The listening study of Section~\ref{sec:listening} provides one human
reference point on the same passages. Listeners preferred \modelname{} over
Fish Audio S2 Pro in 70.25\% of forced choices, while the judge's prosody score
for that comparison is 82.1\%. Both favor \modelname{}, with a larger margin
from the judge. The two evaluations used different \modelname{} renders, and the
judge scores ties as one half, so this agreement in direction is not a
per-pair measure of agreement between the judge and listeners. The comparisons
with ElevenLabs, Gradium, and Cartesia have no human counterpart. We do not
claim that this exact Gemini 3.1 protocol has been independently validated
against human judgments, or that model judges are superior to human raters.

The benchmark measures English reading prosody and word-level correctness. It
does not establish voice similarity, general sound quality, German or broader
multilingual performance, long-form continuity, or streaming quality.

\subsection{Intelligibility}
\label{sec:intelligibility}

On the 1,088 English zero-shot examples of the Seed-TTS evaluation set
\citep{anastassiou2024seedtts}, \modelname{} achieves a mean utterance-level
word error rate (WER) of 1.66\% at semantic temperature 0.6, using Whisper
large-v3 transcription \citep{radford2023whisper,openaiwhisperlargev3}. For
reference, the Seed-TTS paper reports 2.25\% for Seed-TTS and 2.14\% for the
human recordings on a 1,000-utterance version of this set.

\subsection{Long-form generation}
\label{sec:longform}

\paragraph{Protocol.}
We test whether repeated window advancement degrades final-chunk
intelligibility or speaker similarity relative to a short-generation baseline.
We generate each passage either as ten consecutive chunks (about three minutes)
or as only its last two, and score the final chunk. Both conditions use
identical final-chunk text and preceding-chunk text; the preceding audio is
generated separately in each condition. We use 60 PG-19 test passages from
distinct books: ten for protocol development and 50 for evaluation. Each chunk
contains 230--340 characters. Both conditions use the same voice reference and
matched per-passage seeds, with one run at each semantic temperature, 0.7 and
0.8. We apply a frequency penalty of 0.1 within each chunk's semantic stream to
reduce repetitive generation observed in preliminary runs. Text verbalization,
silence trimming, pause shortening, and super-resolution are disabled.

\paragraph{Results.}
On the 50 evaluation passages, mean final-chunk WER is 3.7\% versus 3.9\% for
long versus short generation at temperature 0.7, and 4.8\% versus 3.8\% at
0.8. The mean WER difference (long minus short), paired by passage, is
\(-0.2\) percentage points at temperature 0.7 (95\% bootstrap CI: \(-0.5\) to
\(+0.1\)) and \(+1.0\) at 0.8 (\(-0.2\) to \(+3.3\)). WER uses Whisper
large-v3 on the isolated final chunk. WavLM/ECAPA-TDNN speaker similarity
\citep{chen2022wavlm,desplanques2020ecapa} to the reference decreases by 0.013
at 0.7 and 0.020 at 0.8.

Across all 60 passages, silence-related text omissions occur in two long
generations at each temperature and in no short generations. Only the final
chunk is transcribed; earlier chunks are checked for silence and repetition.
Final-chunk scores therefore do not establish complete delivery.

\subsection{Discussion}

In the blinded study, most listeners preferred the system's pauses, emphasis,
and rhythm over those of Fish Audio S2 Pro. This supports its prosodic quality
on English reading passages. Separate ASR results support its intelligibility. The LLM-as-a-judge benchmark agrees in direction
with listeners for Fish Audio S2 Pro and suggests prosody comparable to
ElevenLabs Flash v2.5, but its margins against the other systems have no human
counterpart.

Final-chunk WER was close to the short-generation baseline at temperature 0.7;
at 0.8, the increase was driven by one severe final-chunk omission. Together
with the modest reduction in speaker similarity, this suggests limited
degradation at the final chunk under the tested settings, although occasional
omissions remain. Some recordings recovered after earlier failures, so endpoint
performance alone does not establish continuity throughout a passage.

Both generation lengths use the same chunking scheme, so these results do not
isolate its contribution or those of its components; that requires controlled
ablations.

\section{Limitations and Release}
\label{sec:limits}

\paragraph{Technical limitations.}
Autoregressive semantic generation can omit, repeat, or alter text and can
terminate too early or too late. Reference conditioning may transfer identity
imperfectly or reproduce incidental recording properties. Long-form chunking
can introduce discontinuities; in our long-form experiment, silence-related
text omissions occurred only in long generations (Section~\ref{sec:longform}).
The serial four-stage factorization adds
latency. The optional generative verbalizer can normalize incorrectly or alter
wording. Accepted language or style labels specify the input contract; they do
not by themselves establish equal quality or complete coverage. The supported
language labels are \texttt{english}, \texttt{german}, \texttt{spanish},
\texttt{french}, \texttt{italian}, \texttt{dutch}, and
\texttt{portuguese}. In informal listening, German intonation is strong but
phoneme realization is sometimes inaccurate; the remaining languages have not
been checked by native speakers, so we report no conclusions about them.
Training was weighted toward audiobook speech, so audiobook generation may be
more reliable than conversational or agentic generation. The system should be evaluated on
the intended domain, language, speakers, text lengths, and deployment hardware
before use.

\paragraph{Safety.}
Voice cloning can enable impersonation, fraud, nonconsensual synthesis, and
misleading media. The release includes synthetic reference voices,
generated by the model rather than recorded from speakers. Users remain
responsible for consent on references they supply. Deployers should obtain
permission for reference voices,
authenticate callers, rate-limit and log access, disclose that generated audio
is synthetic where appropriate, and maintain abuse-response procedures.

\paragraph{Release boundary and attribution.}
The TTS weights are distributed under the Tontaube Community Model License 1.0
included with the Hugging Face model release, which is not an open-source license. The license defines
the permitted uses and applicable commercial requirements; readers should
consult it for the terms that apply. The optional verbalizer and inference
implementation are both distributed separately under the Apache License 2.0.
Users must also comply with the notices and licenses applicable
to Qwen3, DualCodec, VibeVoice, vLLM, and other third-party components used by the implementation.

{\fontsize{9.5pt}{9.7pt}\selectfont
\bibliographystyle{unsrtnat}
\setlength{\bibsep}{0pt}
\bibliography{references}

@misc{qwen3technicalreport,
  title         = {{Qwen3} Technical Report},
  author        = {An Yang and Anfeng Li and Baosong Yang and others},
  year          = {2025},
  eprint        = {2505.09388},
  archiveprefix = {arXiv},
  primaryclass  = {cs.CL},
  url           = {https://arxiv.org/abs/2505.09388}
}

@inproceedings{li2025dualcodec,
  title     = {{DualCodec: A Low-Frame-Rate, Semantically-Enhanced Neural Audio Codec for Speech Generation}},
  author    = {Jiaqi Li and Xiaolong Lin and Zhekai Li and Shixi Huang and Yuancheng Wang and Chaoren Wang and Zhenpeng Zhan and Zhizheng Wu},
  year      = {2025},
  booktitle = {Proceedings of Interspeech 2025},
  pages     = {4883--4887},
  doi       = {10.21437/Interspeech.2025-468},
  url       = {https://www.isca-archive.org/interspeech_2025/li25e_interspeech.html}
}

@misc{dualcodecsoftware,
  title        = {{DualCodec} 0.4.2},
  author       = {Jiaqi Li and others},
  year         = {2025},
  howpublished = {Software release},
  note         = {Released 22 August 2025},
  url          = {https://pypi.org/project/dualcodec/0.4.2/}
}

@inproceedings{peng2026vibevoice,
  title     = {{VibeVoice}: Expressive Podcast Generation with Next-Token Diffusion},
  author    = {Zhiliang Peng and Jianwei Yu and Wenhui Wang and Yaoyao Chang and Yutao Sun and Li Dong and Yi Zhu and Weijiang Xu and Hangbo Bao and Zehua Wang and Shaohan Huang and Yan Xia and Furu Wei},
  booktitle = {International Conference on Learning Representations},
  year      = {2026},
  url       = {https://openreview.net/forum?id=FihSkzyxdv}
}

@misc{microsoftvibevoiceacoustic,
  title        = {{VibeVoice}-1.5{B}},
  author       = {{Microsoft}},
  year         = {2026},
  howpublished = {Hugging Face model revision \texttt{c00898d257e6}},
  url          = {https://huggingface.co/microsoft/VibeVoice-1.5B/tree/c00898d257e6b46004e3e2866a47534085fb685a}
}

@inproceedings{kwon2023pagedattention,
  title     = {Efficient Memory Management for Large Language Model Serving with {PagedAttention}},
  author    = {Woosuk Kwon and Zhuohan Li and Siyuan Zhuang and Ying Sheng and Lianmin Zheng and Cody Hao Yu and Joseph E. Gonzalez and Hao Zhang and Ion Stoica},
  booktitle = {Proceedings of the 29th Symposium on Operating Systems Principles},
  pages     = {611--626},
  year      = {2023},
  publisher = {ACM},
  doi       = {10.1145/3600006.3613165}
}

@article{su2024rope,
  title   = {{RoFormer}: Enhanced Transformer with Rotary Position Embedding},
  author  = {Jianlin Su and Murtadha Ahmed and Yu Lu and Shengfeng Pan and Wen Bo and Yunfeng Liu},
  journal = {Neurocomputing},
  volume  = {568},
  pages   = {127063},
  year    = {2024},
  doi     = {10.1016/j.neucom.2023.127063}
}

@inproceedings{rae2020compressive,
  title     = {Compressive Transformers for Long-Range Sequence Modelling},
  author    = {Jack W. Rae and Anna Potapenko and Siddhant M. Jayakumar and Chloe Hillier and Timothy P. Lillicrap},
  booktitle = {International Conference on Learning Representations},
  year      = {2020},
  url       = {https://openreview.net/forum?id=SylKikSYDH}
}

@inproceedings{manku2025emergentttseval,
  title     = {{EmergentTTS-Eval}: Evaluating {TTS} Models on Complex Prosodic, Expressiveness, and Linguistic Challenges Using Model-as-a-Judge},
  author    = {Ruskin Raj Manku and Yuzhi Tang and Xingjian Shi and Mu Li and Alexander J. Smola},
  booktitle = {Advances in Neural Information Processing Systems},
  volume    = {38},
  pages     = {3514--3564},
  year      = {2025},
  doi       = {10.52202/085713-0110},
  note      = {Datasets and Benchmarks Track}
}

@misc{vllm2026registration,
  title        = {Registering a Model},
  author       = {{vLLM Project}},
  year         = {2026},
  howpublished = {vLLM 0.16.0 documentation},
  url          = {https://docs.vllm.ai/en/v0.16.0/contributing/model/registration/}
}

@misc{vllm2026ngram,
  title        = {{N-Gram} Speculation},
  author       = {{vLLM Project}},
  year         = {2026},
  howpublished = {vLLM 0.16.0 documentation},
  url          = {https://docs.vllm.ai/en/v0.16.0/features/spec_decode/}
}

@misc{fishaudio2026s2pro,
  title        = {{Fish Audio S2 Pro}},
  author       = {{Fish Audio}},
  year         = {2026},
  howpublished = {Hugging Face model revision \texttt{1de9996b6be3}},
  url          = {https://huggingface.co/fishaudio/s2-pro/tree/1de9996b6be38b745688de084d87a5633f714e4e}
}

@misc{elevenlabs2026flash25,
  title        = {{Eleven Flash v2.5}},
  author       = {{ElevenLabs}},
  year         = {2026},
  howpublished = {Official Replicate deployment \texttt{elevenlabs/flash-v2.5}},
  note         = {Evaluated April 2026 with voice \texttt{James}},
  url          = {https://replicate.com/elevenlabs/flash-v2.5/readme}
}

@misc{gradium2026tts,
  title        = {{Gradium Text-to-Speech API}},
  author       = {{Gradium}},
  year         = {2026},
  howpublished = {API model \texttt{default}},
  note         = {Evaluated April 2026 with voice ID \texttt{QETTJoT4n\_WmpL3w}},
  url          = {https://docs.gradium.ai/guides/release-notes}
}

@misc{cartesia2026sonic3,
  title        = {{Cartesia Sonic 3}},
  author       = {{Cartesia}},
  year         = {2026},
  howpublished = {API model alias \texttt{sonic-3}},
  note         = {Evaluated April 2026 with voice ID \texttt{79f8b5fb-2cc8-479a-80df-29f7a7cf1a3e}},
  url          = {https://docs.cartesia.ai/build-with-cartesia/tts-models/older-models}
}

@misc{google2026gemini31pro,
  title        = {{Gemini 3.1 Pro Preview}},
  author       = {{Google}},
  year         = {2026},
  howpublished = {Google Cloud model documentation},
  note         = {Model ID \texttt{gemini-3.1-pro-preview}; accessed 18 August 2026},
  url          = {https://docs.cloud.google.com/vertex-ai/generative-ai/docs/models/gemini/3-1-pro}
}

@misc{liao2024fishspeech,
  title         = {{Fish-Speech}: Leveraging Large Language Models for Advanced Multilingual Text-to-Speech Synthesis},
  author        = {Shijia Liao and Yuxuan Wang and Tianyu Li and Yifan Cheng and Ruoyi Zhang and Rongzhi Zhou and Yijin Xing},
  year          = {2024},
  eprint        = {2411.01156},
  archiveprefix = {arXiv},
  primaryclass  = {cs.SD},
  url           = {https://arxiv.org/abs/2411.01156}
}

@misc{qwen2026tts,
  title         = {{Qwen3-TTS} Technical Report},
  author        = {Hangrui Hu and Xinfa Zhu and Ting He and others},
  year          = {2026},
  eprint        = {2601.15621},
  archiveprefix = {arXiv},
  primaryclass  = {cs.SD},
  url           = {https://arxiv.org/abs/2601.15621}
}

@inproceedings{copet2023musicgen,
  title     = {Simple and Controllable Music Generation},
  author    = {Jade Copet and Felix Kreuk and Itai Gat and Tal Remez and David Kant and Gabriel Synnaeve and Yossi Adi and Alexandre D{\'e}fossez},
  booktitle = {Advances in Neural Information Processing Systems},
  volume    = {36},
  pages     = {47704--47720},
  year      = {2023},
  doi       = {10.52202/075280-2066},
  url       = {https://proceedings.neurips.cc/paper_files/paper/2023/hash/94b472a1842cd7c56dcb125fb2765fbd-Abstract-Conference.html}
}

@misc{lyth2024parler,
  title         = {Natural Language Guidance of High-Fidelity Text-to-Speech with Synthetic Annotations},
  author        = {Dan Lyth and Simon King},
  year          = {2024},
  eprint        = {2402.01912},
  archiveprefix = {arXiv},
  primaryclass  = {cs.SD},
  url           = {https://arxiv.org/abs/2402.01912}
}

@inproceedings{zhang2023speechtokenizer,
  title     = {{SpeechTokenizer}: Unified Speech Tokenizer for Speech Language Models},
  author    = {Xin Zhang and Dong Zhang and Shimin Li and Yaqian Zhou and Xipeng Qiu},
  booktitle = {International Conference on Learning Representations},
  year      = {2024},
  url       = {https://openreview.net/forum?id=AF9Q8Vip84}
}

@inproceedings{chung2021w2vbert,
  title     = {{w2v-BERT}: Combining Contrastive Learning and Masked Language Modeling for Self-Supervised Speech Pre-Training},
  author    = {Yu-An Chung and Yu Zhang and Wei Han and Chung-Cheng Chiu and James Qin and Ruoming Pang and Yonghui Wu},
  booktitle = {2021 IEEE Automatic Speech Recognition and Understanding Workshop},
  pages     = {244--250},
  year      = {2021},
  doi       = {10.1109/ASRU51503.2021.9688253}
}

@misc{defossez2024moshi,
  title         = {Moshi: a speech-text foundation model for real-time dialogue},
  author        = {Alexandre D{\'e}fossez and Laurent Mazar{\'e} and Manu Orsini and Am{\'e}lie Royer and Patrick P{\'e}rez and Herv{\'e} J{\'e}gou and Edouard Grave and Neil Zeghidour},
  year          = {2024},
  eprint        = {2410.00037},
  archiveprefix = {arXiv},
  primaryclass  = {eess.AS},
  url           = {https://arxiv.org/abs/2410.00037}
}

@misc{tontaube2026v0,
  title        = {{TontaubeV0} Model Card},
  author       = {Fritz Cremer and Jonathan Cremer},
  year         = {2026},
  month        = apr,
  howpublished = {Model card, Tontaube},
  url          = {https://tontaube.ai/blog/tontaube-v0-model-card}
}

@misc{openaiwhisperlargev3,
  title        = {{Whisper Large v3}},
  author       = {{OpenAI}},
  year         = {2023},
  howpublished = {Hugging Face model card},
  url          = {https://huggingface.co/openai/whisper-large-v3}
}

@misc{anastassiou2024seedtts,
  title         = {{Seed-TTS}: A Family of High-Quality Versatile Speech Generation Models},
  author        = {Philip Anastassiou and others},
  year          = {2024},
  eprint        = {2406.02430},
  archiveprefix = {arXiv},
  url           = {https://arxiv.org/abs/2406.02430}
}

@misc{liao2026fishaudios2,
  title         = {{Fish Audio S2} Technical Report},
  author        = {Shijia Liao and Yuxuan Wang and Songting Liu and Yifan Cheng and Ruoyi Zhang and Tianyu Li and Shidong Li and Yisheng Zheng and Xingwei Liu and Qingzheng Wang and Zhizhuo Zhou and Jiahua Liu and Xin Chen and Dawei Han},
  howpublished  = {arXiv:2603.08823},
  year          = {2026},
  eprint        = {2603.08823},
  archiveprefix = {arXiv},
  url           = {https://arxiv.org/abs/2603.08823}
}

@inproceedings{radford2023whisper,
  title     = {Robust Speech Recognition via Large-Scale Weak Supervision},
  author    = {Alec Radford and Jong Wook Kim and Tao Xu and Greg Brockman and Christine {McLeavey} and Ilya Sutskever},
  booktitle = {Proc. 40th International Conference on Machine Learning},
  series    = {Proceedings of Machine Learning Research},
  volume    = {202},
  pages     = {28492--28518},
  year      = {2023},
  url       = {https://proceedings.mlr.press/v202/radford23a.html}
}

@inproceedings{zhao2025clearervoice,
  title     = {{ClearerVoice-Studio}: Bridging Advanced Speech Processing Research and Practical Deployment},
  author    = {Shengkui Zhao and Zexu Pan and Bin Ma},
  booktitle = {Proc. Interspeech},
  pages     = {2980--2984},
  year      = {2025},
  doi       = {10.21437/Interspeech.2025-680}
}

@article{chen2022wavlm,
  title   = {{WavLM}: Large-Scale Self-Supervised Pre-Training for Full Stack Speech Processing},
  author  = {Sanyuan Chen and Chengyi Wang and Zhengyang Chen and Yu Wu and Shujie Liu and Zhuo Chen and Jinyu Li and Naoyuki Kanda and Takuya Yoshioka and Xiong Xiao and Jian Wu and Long Zhou and Shuo Ren and Yanmin Qian and Yao Qian and Jian Wu and Michael Zeng and Xiangzhan Yu and Furu Wei},
  journal = {IEEE Journal of Selected Topics in Signal Processing},
  volume  = {16},
  number  = {6},
  pages   = {1505--1518},
  year    = {2022},
  doi     = {10.1109/JSTSP.2022.3188113},
  url     = {https://doi.org/10.1109/JSTSP.2022.3188113}
}

@inproceedings{desplanques2020ecapa,
  title     = {{ECAPA-TDNN}: Emphasized Channel Attention, Propagation and Aggregation in {TDNN} Based Speaker Verification},
  author    = {Brecht Desplanques and Jenthe Thienpondt and Kris Demuynck},
  booktitle = {Proc. Interspeech},
  pages     = {3830--3834},
  year      = {2020},
  doi       = {10.21437/Interspeech.2020-2650},
  url       = {https://www.isca-archive.org/interspeech_2020/desplanques20_interspeech.html}
}
}

\clearpage
\appendix
\section{Chunked generation example}
\label{app:chunked-example}

The example of Section~\ref{sec:io} fits in one chunk. Longer passages are divided into chunks, with
matching boundaries represented in \(\cb{0}\)'s text and audio rows. Consider \emph{Hi
there. How are you? I hope so.} divided into three chunks. When generating the
middle chunk, \(\cb{0}\) receives the following layout:

\begin{center}
\small
\setlength{\extrarowheight}{3pt}
\begin{tabular}{@{}l@{\quad}l@{\qquad}l@{}}
in & \ttfamily <|im\_start|>\sep english\sep \textvisiblespace{}:\sep \textvisiblespace{}audi\sep obook\sep <|im\_end|>\sep \textbackslash n & \\
& \ttfamily H\sep i\sep \textvisiblespace\sep t\sep h\sep e\sep r\sep e\sep . & previous chunk \\
& \ttfamily <|text\_split|> & \\
& \ttfamily \textvisiblespace\sep H\sep o\sep w\sep \textvisiblespace\sep a\sep r\sep e\sep \textvisiblespace\sep y\sep o\sep u\sep ? & current chunk \\
& \ttfamily <|text\_split|> & \\
& \ttfamily \textvisiblespace\sep I\sep \textvisiblespace\sep h\sep o\sep p\sep e\sep \textvisiblespace\sep s\sep o\sep .\sep \textbackslash n & lookahead, up to 50 characters \\
& \ttfamily PAD\sep \(c^0_1\)\sep \(\ldots\)\sep \(c^0_m\) & audio for the previous chunk \\
& \ttfamily <|audio\_split|> & \\
out & \ttfamily \(c^0_{m+1}\)\sep \(\ldots\)\sep \(c^0_L\)\sep <|audio\_split|> & audio for the current chunk only \\
\end{tabular}
\end{center}

\noindent
Four details are visible here that the single-chunk example cannot show. The
current chunk begins with a space, as every chunk but the first does, and does
not end with a newline, which only the last chunk of a passage carries; together
these tell \(\cb{0}\) where it is in the passage. The text row runs past the
current chunk into the beginning of the next, so \(\cb{0}\) receives part of
the next phrase as lookahead before generating the corresponding audio; this
lookahead is truncated at 50
characters, mid-word if necessary. The matching \texttt{<|text\_split|>} and
\texttt{<|audio\_split|>} share a logical position under
Equation~\ref{eq:markers}, so the text and audio rows are realigned at the
boundary even when the text and audio representations of the same chunk contain
different numbers of tokens. And because
this chunk is internal, \(\cb{0}\) stops at \texttt{<|audio\_split|>}
rather than \texttt{<|end\_of\_speech|>}; the window then advances by
discarding the oldest text--audio chunk pair. Only the second
\texttt{<|text\_split|>} has no matching audio marker in the input because the
current chunk's audio is still being generated; the emitted
\texttt{<|audio\_split|>} supplies its counterpart.

\section{Text verbalization}
\label{app:verbalizer}

The optional verbalizer is an independently trained Qwen3-1.7B model that maps
written English to the spoken-form text \(X\) consumed by \modelname{}. It is
instructed to expand numbers, dates, times, currencies, and symbols, to
standardize abbreviations and initialisms for pronunciation, and otherwise to
preserve the wording. Ordinary spans use the Qwen tokenizer, while each digit in
a numeric sequence is encoded separately. Inference runs at temperature zero
with vLLM's n-gram prompt-lookup speculative decoding~\citep{vllm2026ngram}.

The checkpoint is English-only, is independent of the four-stage graph, and is
excluded from Table~\ref{tab:parameters}. Because it is generative, it can
normalize incorrectly or alter wording; callers that need exact control can
bypass it and supply spoken-form text directly.

\clearpage
\section{LLM-as-a-Judge Instructions}
\label{app:judge-prompt}

For transparency, the instructions supplied with each audio pair are reproduced
below. The placeholder \texttt{\{text\}} was replaced verbatim by the passage
for that row; line wrapping below is typographic.

\smallskip
\hrule height 0.4pt
\smallskip
\begingroup
\small\rmfamily\raggedright
You will hear two text-to-speech audios for the same reference text. Your job
is to compare them on TWO dimensions and pick which is better on each.

\smallskip
REFERENCE TEXT:\\
"\{text\}"

\smallskip
For EACH dimension, answer "FIRST" if the first audio is meaningfully better,
"SECOND" if the second is meaningfully better, or "TIE" if they're basically
equivalent on that dimension.

\smallskip
Dimensions:\\
\mbox{ - }prosody: rhythm, intonation, emphasis, pacing, naturalness\\
\parbox[t]{\linewidth}{\mbox{ - }correctness: did it speak the reference text
accurately (word-by-word)}

\smallskip
CRITICAL --- what to IGNORE when comparing:\\
\mbox{ - }Audio / sound quality (hiss, compression, noise, clipping, tinny
timbre).\\
\mbox{ - }The specific voice timbre (whether it sounds like one speaker vs
another).\\
\mbox{ - }Recording artifacts from the TTS model or codec.

\smallskip
Additional rules:\\
\mbox{ - }Judge each dimension INDEPENDENTLY. A win on prosody does not imply a
win on correctness --- they are separate qualities.\\
\mbox{ - }Do not let audio order (which was played first) influence your
judgment.

\smallskip
Respond with ONLY a raw JSON object on a single line, no markdown, no
commentary:\\
\{"prosody": "FIRST"\textbar"SECOND"\textbar"TIE", "correctness":
"FIRST"\textbar"SECOND"\textbar"TIE",
\mbox{"notes": "\textless one short sentence\textgreater"}\}
\par
\endgroup
\smallskip
\hrule height 0.4pt

\end{document}